    \newcommand{\be}{\begin{equation}} \newcommand{\ee}{\end{equation}}
   \newcommand{\bea}{\begin{eqnarray}}\newcommand{\eea}{\end{eqnarray}}
   \newcommand{\nn}{\nonumber}
   \newcommand\T{\mbox{Tr}}
   \newcommand{\s}{\scriptscriptstyle}
   \documentstyle[12pt]{article}
\begin{document}
   \renewcommand{\thefootnote}{\fnsymbol{footnote}}
   \def\theequation{\arabic{section}.\arabic{equation}}

   \begin{titlepage}

   \hfill hep-th/9512035 ,\hspace{2mm}JINR-E2-95-458

   \vspace{2cm}

   \begin{center}
 {\bf ZERO-CURVATURE REPRESENTATION FOR  HARMONIC-SUPERSPACE EQUATIONS
  OF MOTION IN $ N=1, D = 6$ SUPERSYMMETRIC GAUGE THEORY}\\
   \vspace{1cm}

   {\bf B.M.Zupnik}${}\;$\footnote{E-mail: zupnik@thsun1.jinr.dubna.su}\\
   {\it Bogoliubov   Laboratory of Theoretical Physics , JINR, Dubna,
   Moscow Region, 141980, Russia }\\
   \end{center}

   \vspace{3cm}
   \begin{abstract}
We consider the $SYM^1_6$ harmonic-superspace system of equations that
contains superfield constraints and  equations of motion for the simplest
six-dimensional supersymmetric gauge theory. A special $A$-frame of the
analytic basis is introduced where a kinematic  equation for  the harmonic
connection $A^{\s--}$ can be solved . A dynamical equation in this frame
is equivalent to the zero-curvature equation corresponding to the
covariant conservation of analyticity. Using a simple harmonic  gauge
condition for the gauge group $SU(2)$ we derive the superfield equations
that produce the general $SYM^1_6$ solution . An analogous approach for
the analysis of integrability conditions for the $SYM^2_4$-theory and
$SYM$-supergravity-matter systems in harmonic superspace is  discussed
briefly.
   \end{abstract}
   \end{titlepage}

\renewcommand{\thefootnote}{\arabic{footnote}}
\setcounter{footnote}{0}
\setcounter{section}{0}
\section{Introduction}

$\;\;\;$ Harmonic superspace ($HS$) was introduced in Refs\cite{a1,a2}
in order to consistently describe supergravity, supersymmetric gauge and
matter theories with $N=2, D=4$ supersymmetry. The harmonic approach is a
covariant version of the twistor method and it has been intensively used
for the construction of self-dual solutions in ordinary and supersymmetric
Yang-Mills and gravity theories \cite{a3,a4}.

We shall use the standard notation $SYM^N_D$ for supersymmetric gauge
theories with $(D,N)$-supersymmetry in the space-time of dimension $D$.

It should be noted that  twistor-harmonic  methods have been applied
for the integrability interpretation of the non-self-dual theories
$SYM^3_4$ \cite{a5}-\cite{a10}, $SYM^4_5$ \cite{So} and $SYM^1_{10}$
\cite{a11,a14}, however, the corresponding harmonic superspaces are very
complicated. The simple harmonic $SU(2)/U(1)$ formalism allows the
construction of a general solution to the 3-dimensional $SYM^6_3$
 equations  \cite{a12}.

The notion of {\it $HS$-integrability} is connected with a reformulation
 of the $SYM$-equations  as  conditions of zero-harmonic-superfield
 curvatures constructed by means of  covariant harmonic
derivatives and  harmonized spinor or vector covariant derivatives.
 These conditions can be interpreted as the covariant conservation of
 harmonic analyticity \cite{a1}. The harmonic coordinates in the
superfield formalism of $HS$-theories are the analogues of the auxiliary
 (spectral) parameters. The final construction of $HS$-solutions can be
 reformulated in terms of the ordinary coordinates.

We propose the $HS$-integrability interpretation of the supersymmetric
$N=1, D=6$ gauge theory $SYM^1_6$ connected via a dimensional reduction
with the $SYM^2_4$-theory. The $HS$-formalism of $SYM^1_6$ has been
 considered in Refs\cite{a14,a13,a15} by analogy with \cite{a1}.
A review of the standard harmonic approach is presented in Section
\ref{A}. We call this version of the harmonic formalism {\it the $V$-frame
 of the analytic basis}, because it uses the analytic prepotential
 $V^{\s++}$ \cite{a1} as a basic field variable. Section \ref{B} contains
a new version of the harmonic formalism ($A$-frame) using the nonanalytic
harmonic connection $A^{\s--}$ as an independent variable. It will be
shown that the $SYM^1_6$ superfield constraints and equations of motion
 can be reformulated as an dynamical zero-curvature equation plus a
 linear solvable constraint in this frame. The $HS$-approach produces
 also an infinite number of conservation laws and equations for the
 B\"acklund-transformation matrix  in $SYM^1_6$. Section \ref{D} is
devoted to the analysis of $SYM^1_6$-solutions in the $V$-frame for the
 gauge group $SU(2)$. We use a special harmonic representation of the
 $SU(2)$-prepotential and the simplest harmonic gauge \cite{a12}. This
 gauge simplifies the study of the spontaneously broken phase of
$SYM^1_6$. The $A$-frame analysis of $SU(2)$-solutions is considered in
Section \ref{E}.

In the conclusion the $HS$-integrability of the $SYM^2_4$-theory is
discussed briefly. In particular, self-dual and anti-self-dual solutions
 and a duality transformation have  simple representations in the
$A$-frame. We consider also a possible modifications of the superfield
 $SYM^2_4$-action. We hope that the  $HS$-integrability  of $SYM^2_4$
 can help to understand the remarkable quantum properties of this theory
 \cite{SW}.

This work has been reported on the Workshop "Supersymmetries and Quantum
Symmetries" , Dubna, 1995 and in the short version in Ref\cite{Zup}.

\setcounter{equation}{0}
\section{ \label{A} Harmonic formalism  of
 $SYM^1_6$ }

\hspace{0.5cm}Several versions of superfield formalism produce
 manifestly supersymmetric
descriptions of the off-shell $SYM^1_6$-theory \cite{a14,a15,a16,a17}.
Consider  $N=1, D=6$ superspace $M(6, 8)$ \cite{a16,a17} with the 6 even
vector coordinates $x^{ab}$ and 8 odd spinor coordinates $\theta^a_i$
 where $a, b\ldots$ are  4-spinor indices of
the Lorentz group $\;\;SU^* (4) \sim SO(5,1)\;\;$ and $\;\;i,k\ldots\;\;$
 are
 2-spinor indices of $SU(2)$ group. Let $z = (x^{ab},\theta^a_i )$ be the
short notation for the coordinates in this superspace.

The (plane) spinor derivatives  in $M(6, 8)$ satisfy the
basic relation
\be
\{ D^k_a , D^l_b \}=i\varepsilon^{kl} \partial_{ab}   \label{A1}
\ee
where $\partial_{ab}=\partial /\partial x^{ab}$. We shall use the
following combinations of the spinor derivatives \cite{a16}
\bea
& (D_2)_{ab}=(1/2)\varepsilon_{ik}\;D^i_{(a}\;D^k_{b)} &\\ \label{A2}
& (D_4)^{iklm}=(1/24)\varepsilon^{abcd}\;D^i_a\;D^k_b\;D^l_c\;D^m_d &
\label{A3}
\eea
where parentheses denote  symmetrization of the indices.
These satisfy  the useful identities  \cite{a16}:
\be
D^{(n}_a (D_4)^{iklm)}=0,\hspace{1cm}(D_2)_{ab}(D_4)^{iklm}=0
\label{A3b}
\ee

By analogy with the $SYM^2_4$-theory \cite{GSW} the superfield constraints
 of $SYM^1_6$  can be written in the following form
\cite{a16,a17}:
\be
\{ \nabla^i_a , \nabla^k_b \} + \{ \nabla^k_a , \nabla^i_b \}= 0
\label{A4}
\ee
where $\nabla^i_a= D^i_a + A^i_a (z)$ is the covariant spinor derivative
and $A^i_a$ is the spinor gauge superfield in a central basis $(CB)$.

The $SYM^1_6$ superfield equation of motion has  dimension $d=-2$ in
 units of length
\be
\nabla^i_a W^{ak} + \nabla^k_a W^{ai}= 0
\label{A5}
\ee
where $ W^{ai}$ is the covariant superfield-strength  of $SYM^1_6$
\be
 W^{ak}=(i/12)
 \varepsilon^{abcd}\varepsilon_{jl}[\nabla^k_b\;,\{\nabla^j_c\;,
\nabla^l_d\}]
\label{A5b}
\ee

The integrable superfield constraints (\ref{A4}) can be solved in the
harmonic approach, and this solution generates a covariant off-shell
 description of the $SYM^1_6$ theory.
  The integrability of the
 whole $SYM^1_6$-system  including Eqs (\ref{A4}) and (\ref{A5}) will be
discussed below.

We shall use the standard   notation
for $SU(2)/U(1)$ harmonics $u^\pm_i$ and the partial harmonic derivatives
\bea
& [\partial^{\s++},\partial^{\s--}]=\partial^{0} &\\ \nn
&  \partial^{\s++}\;u^+_i=0,\;\;\; \partial^{\s++}\;u^-_i=u^+_i &\\
 \label{A6}
&  \partial^{\s--}\;u^-_i=0,\;\;\; \partial^{\s--}\;u^+_i=u^-_i &\nn
\eea
where $\partial^{0}$ is the operator corresponding to $U(1)$-charge $q$.
These harmonics and derivatives have simple representations in terms of
the real $U(1)$-variable $\varphi$ and the complex spectral variable
 $\lambda$ (see e.g.\cite{a1,a19})
\be
\left(\begin{array}{cc} u^-_1 & u^+_1 \\
			u^-_2 & u^+_2 \end{array}\right)=
\frac{1}{\eta(\lambda)}
\left(\begin{array}{cc} 1 &-\lambda\\
		  \bar{\lambda}& 1 \end{array} \right)
\left(\begin{array}{cc} e^{-i\varphi} &0\\
		  0  & e^{i\varphi} \end{array} \right)
\label{A6b}
\ee
where $\eta(\lambda)=\sqrt{1+\lambda\bar{\lambda}}$. The convenient
representation of the harmonic derivatives has the following form:
\be
\partial^{\s++}=e^{ 2i\varphi}
[\eta^2(\lambda)\bar{\partial}_{\lambda} - (i/2)\lambda\partial_\varphi ]
\label{A6c}
\ee
\be
\partial^{\s--}=\;-\;e^{-2i\varphi}
[\eta^2(\lambda)\partial_{\lambda} + (i/2)\bar{\lambda}
\partial_\varphi ],\hspace{0.8cm}\partial^0 =-i\partial_\varphi
\label{A6d}
\ee
where corresponding partial derivatives are introduced.

Consider the harmonic (twistor) transform from the central basis of
$SYM^1_6$ to the analytic basis $(AB)$
\be
u^+_i \nabla^i_a = u^+_i h^{-1}D^i_a h  \label{A7}
\ee
where $h(z,u)$ is a bridge matrix satisfying the basic harmonic equation
 \cite{a1}
\be
(\partial^{\s++} + V^{\s++}) h(z,u)=\nabla^{\s++} h =0 \label{A8}
\ee

We now discuss briefly the terminology of the harmonic approach used in
 this paper. The notion of {\it the basis} ($CB$ or $AB$) includes the
choice of the gauge group representation ($\tau$-group or $\Lambda$-group
\cite{a1}) and the complete set of relations between covariant
derivatives. We use also the notion of {\it the frame} in the analytic
 basis. This means the choice of independent field variables and basic
equations generating the complete system of equations.

The analytic connection $V^{\s++}$ with $q=+2$ (prepotential) determines
the off-shell structure of $SYM^1_6$  in the $V$-frame. It 'lives' in an
 analytic harmonic superspace with the coordinates $\zeta=(x_{\s A}^{ab},
\;\theta^c_+ )$
\be
 x_{{\s A}}^{ab}=x^{ab} + (i/4)(\theta^a_+ \theta^b_-\;-\; \theta^b_+
\theta^a_- )
  \label{A8a}
\ee
\be
 \theta^a_+ =u^i_+ \theta^a_i ,\;\;\;\theta^a_-=u^i_- \theta^a_i
\label{A8b}
\ee

The  differential operators  in the analytic coordinates ($\zeta,\;
\theta^a_- $) have the following form:
\bea
& \partial^{\s++}_{{\s A}}=\partial^{\s++} + (i/2)\theta^a_+ \theta^b_+
\partial_{ab} + \theta^a_+ \partial_a^+ & \\ \label{A9}
&\partial^{\s--}_{{\s A}}=\partial^{\s--} + (i/2)\theta^a_-\theta^b_-
\partial_{ab} + \theta^a_- \partial_a^- & \\ \label{A10}
& D^+_a =\partial_a^+ =\partial /\partial \theta^a_- =u_i^+\;D^i_a &\\
 \label{A11}
& D^-_a =-\partial_a^-\; -i\theta^b_-
\partial_{ab}=u_i^-\;D^i_a & \label{A12}
\eea

Note that the $AB$-superfields can be described in terms of the central
coordinates $z,u$, too. We have the useful relations and definitions for
spinor derivatives:
\be
\{\;D^-_a,D^+_b\;\}=i\partial_{ab},\;\;\;\;(D_2)_{ab}=D^+_{(a}\;D^-_{b)}
\label{A12a}
\ee
\be
(D^+ )^4=u^+_i u^+_k u^+_l u^+_m (D_4)^{iklm}=(1/24)\varepsilon^{abcd}\;
D^+_a\;D^+_b\;D^+_c\;D^+_d \label{A12b}
\ee
\be
(D^{\s+3})^a =(1/6)\varepsilon^{abcd}\;D^+_b\;D^+_c\;D^+_d \label{A12c}
\ee

The $V$-system of equations in the $SYM^1_6$-theory contains  off-shell
 constraints and an equation of motion. The basic $V$-frame constraints
 are:

1) The harmonic zero-curvature ($HZC$) equation   \cite{a14,a18}
\be
[\nabla^{\s++},\;\nabla^{\s--}] - \partial^0 =\partial^{\s++}_A A^{\s--} -
\partial^{\s--}_A V^{\s++} + [V^{\s++} , A^{\s--}] = 0
\label{A13}
\ee
where $A^{\s--}(\zeta,\;\theta^a_-,u)$ is the  harmonic connection
with $q=-2$. The harmonic connections  can be expressed via the bridge
 matrix $h$ but we  treat the $HZC$-equation as an independent basic
equation. The general perturbative and non-perturbative solutions of the
 basic harmonic equations (\ref{A8}) and (\ref{A13}) have been discussed
 in Refs\cite{a2,a14,a12,a18}.

2) The 'kinematic' $V$-analyticity condition ( $VZC$-equation ) [1]
\be
[\nabla^+_a ,\;\nabla^{\s++} ]= D^+_a V^{\s++} =0 \label{A14}
\ee

3) The conventional spinor constraint \cite{a14,a18}
\be
[\nabla^{\s--},\;\nabla^+_a  ] =\nabla^-_a =D^-_a + A^-_a=
D^-_a - D^+_a A^{\s--}
\label{A15}
\ee
This constraint allows us to write the spinor connection $A^-_a$ in terms
of the harmonic connection $A^{\s--}$.

4) The initial $CB$-integrability condition is solved trivially by the
 transition to $AB$ \cite{a1}
\be
u^+_i u^+_k\{ \nabla^i_a , \nabla^k_b \} =0\;\Rightarrow\;
\left\{ \nabla^+_a,\;\nabla^+_b \right\}=\left\{ D^+_a,\;D^+_b \right\}= 0
 \label{A16}
\ee

 Secondary constraints follow from the basic constraints 1)-4)
\be
 [\nabla^{\s--},\;\nabla^-_a  ] =0,\;\;\;\{ \nabla^-_a,\;\nabla^-_b \}=0
\label{A17}
\ee
\be
\{ \nabla^+_a,\;\nabla^-_b \}+\{ \nabla^-_a,\;\nabla^+_b \} =0
\label{A18}
\ee

The $V$-frame $SYM^1_6$-equation of motion has been obtained in Ref
\cite{a14} by the use of a corresponding nonpolynomial action
\be
F^{\s++}=(1/4)D^+_a\;W^{a+}(V)=(D^+ )^4 A^{\s--}(V)=0
\label{A19}
\ee
A perturbative solution for $A^{\s--}$ has the following form
\be
 A^{\s--}(V)=\sum\limits^{\infty}_{n=1} (-1)^n \int du_1
\ldots
du_n \frac{V^{\s++}(z,u_1 )\ldots V^{\s++}(z,u_n )}{(u^+ u^+_1)\ldots
 (u^+_n u^+ )}  \label{A20}
\ee
where the harmonic distributions $(u^+_1 u^+_2 )^{-1}$
 \cite{a2} are used.
 Eq(\ref{A19}) is equivalent to the analyticity condition on the
$AB$-superfield-strength $W^{a+}(V)$.

The harmonic distributions have a simple complex representation
\be
\frac{1}{(u^+_1 u^+_2 )}=e^{-i(\varphi_1+\varphi_2)}
\frac{\eta(\lambda_{\s 1})\;\eta(\lambda_{\s 2})}{\lambda_{\s 1}-
\lambda_{\s 2}}
 \label{A20b}
\ee
Using this representation, Eq(\ref{A6c}) and the known formula for complex
distributions
\be
\frac{\partial}{\partial\bar{\lambda}_{\s 1}}\;\frac{1}{\lambda_{\s 1}-
\lambda_{\s 2}}=\pi \delta (\lambda_{\s 1}-\lambda_{\s 2}) \label{A20c}
\ee
 one can reproduce the
 differential relation \cite{a2}
\be
\partial^{\s++}_1 \frac{1}{(u^+_1 u^+_2 )} =\delta^{(1,-1)}(u_1 ,u_2 )=
\pi e^{i(\varphi_1-\varphi_2)}
\eta^4(\lambda_{\s 1})\delta (\lambda_{\s 1}-\lambda_{\s 2})
\label{A20a}
\ee

The equation of motion simplifies in the  normal $V$-gauge \cite{a1} which
is an analogue of the $WZ$-gauge of the simplest superfield theories. The
prepotential of the normal gauge $V^{\s++}_N$ is nilpotent and does not
 contain pure gauge harmonic component fields
\be
 V^{\s++}_N = (1/2)\theta^a_+ \theta^b_+ A_{ab}(x_{{\s A}}) +
(\theta^{\s+3})_a u^-_i\psi^a_i (x_{{\s A}}) + (\theta_+ )^4\;u^-_i u^-_k
 D^{ik}(x_{{\s A}} )                            \label{A21}
\ee
\[
(\theta^{\s+3} )_a =(1/6)\varepsilon_{abcd} \theta^b_+ \theta^c_+
\theta^d_+ ,\;\;\;\;(\theta_+ )^4 =(1/24)\varepsilon_{abcd}\theta^a_+
\theta^b_+ \theta^c_+ \theta^d_+
\]
where $A_{ab}, \psi^a_i $ and $D^{ik}$ are the component fields of a gauge
supermultiplet.

 $SYM^1_6$-action in the normal gauge is the 4-th order polynomial
\be
 S_N =\sum\limits^{4}_{n=2} \int d^{12}z du_1 \ldots du_n
\frac{\T V_N^{\s++}(z,u_1)\ldots V_N^{\s++}(z,u_n)}{(u_1^+ u_2^+)
\ldots (u_n^+u_1^+)} \label{A22}
 \ee

This action generates the superfield $V$-equation of motion equivalent to
the component $SYM^1_6$ equations of motion. An analysis of the nonlinear
 equation (\ref{A19}) is a difficult problem even in the normal gauge.
 Thus, the $V$-frame is useful for the solution of the off-shell
constraints (\ref{A4}) and quantization \cite{a2} but is not very
convenient for the search of the classical solutions.

The original works on harmonic superspaces \cite{a1,a2} and Refs
\cite{a14,a12} use {\it the regular harmonic functions} $V^{\s++}$ and
$h(z,u)$ treated as the convergent or formal harmonic series.
 Regular harmonic functions $f(u)$ correspond to globally defined
 functions on the sphere $S^2$, and irregular functions can contain
 poles and other singularities. The assumption of regularity
is natural for the perturbation theory (e.g. in the normal gauge) but it
leads to unreasonable restrictions on the nonperturbative solutions. We
 shall discuss the irregular bridge functions in Section \ref{B}.

\setcounter{equation}{0}
\section{\label{B} New harmonic frame for the $SYM^1_6$  equations
 of motion}

$\;\;\;$Now we shall consider a new harmonic representation of the
$SYM^1_6$ equations  which allows  to prove the $HS$-integrability of
 this theory and to solve the equation with a dimension $d=-2$.
 Only the complete system of covariant equations in the analytic basis
has an invariant meaning, however, one can change the choice of field
variables and  independent equations. A basic field variable of the
$V$-frame is $V^{\s++}=h\partial^{\s++}h^{-1}$. It is clear that one
can use other functions of the bridge $h$ as the field variables of $AB$.

Let us treat the harmonic connection $A^{\s--}=h\partial^{\s--}h^{-1}$ as
 a basic  superfield of the classical $SYM^1_6$ theory in {\it the
$A$-frame}. The complete $A$-system of $SYM^1_6$-equations for covariant
derivatives with $d\ge -2$ is identical to the corresponding $V$-system,
 however we change {\it the interpretation, the basic set and the order of
 the dynamical equations and the auxiliary field structure} of
 the harmonic formalism in the new frame. The $HZC$-equation (\ref{A13})
 in this frame is treated as an integrable equation for the connection
 $V^{\s++}(A^{\s--})$. A basic $A$-bridge equation contains the covariant
derivative $\nabla^{\s--}$, and Eq(\ref{A8}) becomes a secondary equation.
  Harmonic equations with  $d=0$ do not guarantee the conservation of
 analyticity. We shall preserve the standard transform between $CB$ and
$AB$ (\ref{A7}), the basic $AB$ constraints (\ref{A13}),(\ref{A15}) and
(\ref{A16}) but treat analyticity in the $A$-frame as a new dynamic
 zero-curvature equation  instead of the 'kinematic' analyticity
 constraint of the $V$-frame (\ref{A14}).

It should be underlined that the nonlinear in $V^{\s++}$ equation
(\ref{A19}) transforms to a linear kinematic constraint of the new
 $A$-frame:
\be
(D^+ )^4 A^{\s--}(z,u)=0 \label{B1}
\ee
Using a nilpotency of $D^+_a$ we can obtain the following general solution
of this constraint:
\be
 A^{\s--}(z,u) = D^+_a\;A^{a\s(-3)}(z,u)
\label{B2}
\ee
where $A^{a\s(-3)}$ is the on-shell $SYM^1_6$ prepotential.

Now the whole $SYM^1_6$-system reduces to the dynamic analyticity
(zero-spinor-curvature) condition which we shall call $AZC$-equation
\be
[ \nabla^-_a ,\;\nabla^{\s--} ] =D^-_a\;A^{\s--} + \partial^{\s--}\;D^+_a
\;A^{\s--} - [D^+_a\;A^{\s--},\;A^{\s--}]=0 \label{B3}
\ee
where  the constraint (\ref{A15}) is used.

 This condition and the representation (\ref{B2}) generate
a nonlinear equation for the superfield $A^{a\s(-3)}$
\be
D^-_a\;D^+_b\;A^{b\s(-3)} + \partial^{\s--}\;D^+_a\;D^+_b\;A^{b\s(-3)}
- [D^+_a\;D^+_b\;A^{b\s(-3)},\;D^+_c\;A^{c\s(-3)}]=0 \label{B4}
\ee

This equation has the following gauge invariance:
\be
\delta A^{a\s(-3)}= R^{a\s(-3)}\Lambda + [\Lambda, A^{a\s(-3)}] + D^+_b
\Lambda^{ab\s(-4)} \label{B5}
\ee
where a general symmetrical spinor $\Lambda^{ab\s(-4)}$ and an analytic
scalar $\Lambda$ are the Lie-algebra valued superfield gauge parameters.
The spinor derivative of $\delta A^{a\s(-3)}$ produces the standard
$AB$-gauge transformation $\delta A^{\s--} = \nabla^{\s--}\Lambda$
\be
 \left\{D^+_a , R^{a\s(-3)}\right\}=\partial^{\s--}_A \label{B6}
\ee
\be
R^{a\s(-3)}=\theta^a_-\partial^{\s--} + \frac{i}{4}\theta^a_-\theta^b_-
\theta^c_-\partial_{bc} + \frac{1}{2}\theta^a_-\theta^b_-\partial_b^-
\label{B7}
\ee

Let us consider a regular harmonic functions $A^{a\s(-3)}(z,u)$ and
choose a normal $A$-gauge for the on-shell superfield $A^{\s--}=
D^+_a A^{a\s(-3)}$
\be
A^{\s--}_N =\theta^a_-\beta_a^-(\zeta,u) +\frac{1}{2}\theta^a_-\theta^b_-
\alpha_{ab}(\zeta,u) + (\theta^{-3})_a \psi^{a+}(\zeta,u) \label{B8}
\ee
\[
(\theta^{\s-3})_a = D^+_a (\theta^-)^4 =(1/6) \varepsilon_{abcd}
\theta^b_-\theta^c_-\theta^d_-
\]
where $\beta,\;\alpha$ and $\psi$ are the analytic functions. This gauge
has a residual gauge invariance with restricted parameters
$\partial^{\s--}\Lambda=0$
\be
\delta\beta_a^- = \partial^-_a \Lambda + \ldots,\;\;\;
\delta\alpha_{ab} = \partial_{ab} \Lambda + \ldots \label{B9}
\ee
Note that $(\theta_-)^4$ term vanishes due to the constraint (\ref{B2}).

The superfield $A^{\s--}_N$ contains a physical part $A^{\s--}_P$ and
an auxiliary-field part $H^{\s--}$. All auxiliary harmonic component
 fields vanish as a consequence of Eq(\ref{B4}) so the physical harmonic
 connection be
\be
A^{\s--}_P = (1/2)\theta^a_-\theta^b_- [A_{ab} +
\varepsilon_{abcd}\theta^c_+ u^i_-\psi^d_i]
 + (\theta^{\s-3})_a [u^i_+ \psi^a_i  + \theta^b_+ F^a_b] \label{B10}
\ee
where $A_{ab}(x_{{\s A}})$ and $\psi^a_i(x_{{\s A}}) $ are the vector and
 spinor fields and $F^a_b(x_{{\s A}})$ is an independent field-strength.

Eq(\ref{B4}) generates the usual connection between $F^a_b$ and $A_{ab}$
 and the component $SYM^1_6$ equations. Thus, the $A$-frame corresponds
 to the first-order component $SYM^1_6$ formalism. It is evident that all
 frames of $AB$ are equivalent on-shell and have identical component
 solutions for the physical fields.

 It should be underlined that an alternative equivalent form of the
 $HS$-integrability condition in the $A$-frame can be written as a
 dynamical $VZC$-equation
\be
[\nabla^+_a ,\;\nabla^{\s++} ]=D^+_a\;V^{\s++}(A^{\s--}) = 0
 \label{B10a}
\ee
where  $V^{\s++}(A^{\s--})$ is a solution of Eq(\ref{A13}). A perturbative
form of this solution is an analogue of the solution (\ref{A20}) but
contains the new harmonic distribution
 \be
\frac{1}{(u^-_1 u^-_2 )}=e^{i(\varphi_1+\varphi_2)}
\frac{\eta(\lambda_{\s 1})\;\eta(\lambda_{\s 2})}
{\bar{\lambda}_{\s 1}-\bar{\lambda}_{\s 2}}
 \label{B10b}
\ee
 satisfying the relation
\be
\partial^{\s--}_1 \frac{1}{(u^-_1 u^-_2 )} =\delta^{(-1,1)}(u_1 ,u_2 )
\label{B12}
\ee

The third equivalent form of the dynamical $A$-frame equation can be
written as
\be
[\nabla^{\s++},\nabla^-_a]=\nabla^+_a \Rightarrow [\nabla^{\s++},
[\nabla^{\s--},\nabla^-_a]]=0 \label{B12b}
\ee

The bridge matrix $h_{\s A}=h(A^{\s--})$ of the $A$-frame is a solution
of the following harmonic equation
\be
\nabla^{\s--}\;h_{\s A} = (\partial^{\s--} + D^+_a\;A^{a\s(-3)}\;)\;
h_{\s A} = 0
\label{B13}
\ee
This equation on the sphere $SU(2)/U(1)$ is the harmonic part of the
linear problem for the $HS$-integrable $SYM^1_6$-system. A key point of
the harmonic approach is the integrability of the bridge harmonic
 equation. If we restrict ourselves by regular solutions for $h_{\s A}$,
 then the consistency conditions on the regular harmonic connections
  appear \cite{Iv,a12}. The explicit solutions for the $SU(2)$ gauge group
   be considered in sections \ref{D} and \ref{E}.

Consider a typical example of the linear harmonic differential equation
that arises in an analysis of the bridge equation
\be
\partial^{\s++} f^{\s(-2)}=f^0 (u)=c + c^{(ik)} u^+_i u^-_k +\ldots
\label{B14}
\ee
where $f^0$ is a regular harmonic function. For $c\ne 0$ this equation has
no regular solution in terms of the harmonic expansion. However,
Eq(\ref{B14}) in the complex coordinates (\ref{A6b})
has the integral solution with a simple complex pole kernel. The harmonic
 analogue of this integral representation is
\be
f^{\s(-2)}(u)=\int du_{\s 1} G^{\s(-2,0)}(u,u_{\s 1}) f^0(u_{\s 1})
\label{B15a}
\ee
\be
\partial^{\s++}G^{\s(-2,0)}(u,u_{\s 1})=\delta^{\s(0,0)}(u,u_{\s 1})=
\pi \eta^4(\lambda)\delta (\lambda -\lambda_{\s 1})
  \label{B15}
\ee

In contrast to the standard harmonic distributions \cite{a2}
$G^{\s(-2,0)}(u,u_{\s 1})$ has not any illustrative harmonic expansion,
but  it has the simple complex representation
\be
G^{\s(-2,0)}
(\lambda,\lambda_{\s 1})=e^{-2i\varphi} \frac{\eta^2(\lambda)}
{(\lambda-\lambda_{\s 1})}
\ee

Thus, one can admit the appearance of isolated harmonic singularities
 in the bridge function and even in the harmonic connections. As a rule
we shall use regular initial data and choose the gauge freedom to
obtain the solutions with a minimal number of singularities.

 Using irregular harmonic fields one should remember the following simple
general rule \footnote{The importance of this rule for the harmonic method
was remarked  by V.I.Ogievetsky}: {\it The physical component fields are
 defined naturally in the central basis}.

The $CB$-gauge superfield does not depend on the harmonics
\be
A^i_a(z)=h^{-1}\;D^i_a h - u^{+i}h^{-1}\;(D^+_a A^{\s--})\;h
\label{B16}
\ee
This superfield satisfies the relations $\partial^{\s\pm\pm}A^i_a(z)=0$
and also the equations (\ref{A4}) and (\ref{A5}) which are
 equivalent to the component $SYM^1_6$-equations.

\setcounter{equation}{0}
\section{\label{C} Conservation laws and B\"acklund
transformations in $SYM^1_6$}

$\;\;\;$ The most attractive feature of  integrable field
theories is  an infinite number of conservation laws. The explicit
construction of the conserved quantities follows immediately from
the zero-curvature representation and has a clear geometric interpretation
in terms of the contour variables \cite{a23}. Analogous constructions
arise also for the  integrable $SYM^3_4$ equation \cite{a9}.

The $HS$-integrable theories possess the specific properties. The
 corresponding zero-curvature equations contain covariant spinor and
 harmonic derivatives and mean a conservation of the analyticity in $HS$
 \cite{a1}. Now we shall try to show that ordinary conservation laws
 follow from the dynamic harmonic-spinor analyticity equation of
$SYM^1_6$-theory. Consider a vector covariant derivative in the $A$-frame
\be
\nabla_{ab} = -i\left\{\nabla^+_a,\;\nabla^-_b\right\} = \partial_{ab} +
i D^+_a\;D^+_b\;A^{\s--} \label{C1}
\ee

The basic equation (\ref{B4}) generates the relation
\be
[\nabla^{\s--},\;\nabla_{ab}] = 0 \label{C2}
\ee

Let us choose a time variable $t=x^{12}$
\be
\nabla_t = \nabla_{12} = \partial_t + A_{12},\;\;\;\;A_{12}(z,u)=i D^+_1\;
D^+_2\;A^{\s--} \label{C3}
\ee
It is evident that $\nabla_t$ commutes on-shell with the  covariant
harmonic derivatives (\ref{C2}).

It should be stressed that the bridge  is a natural harmonic
 analogue of the  contour variables of  integrable theories in the
zero-curvature representation . The transformation law of the bridge has
the following form:
\be
\delta h_{\s A} = \Lambda(\zeta,u)\;h_{\s A} - h_{\s A}\;\tau(z)
\label{C5}
\ee
where $\Lambda$ and $\tau$ are the gauge parameters in $AB$ and $ CB$,
correspondingly. The covariant constancy of the bridge in the all spinor
and vector directions is a consistency condition for the dynamic
analyticity equations (\ref{B4}) or (\ref{B10a}), for instance
\be
\nabla_t\;h_{\s A} = \partial_t\;h_{\s A} + A_{12}\;h_{\s A} -
h_{\s A} \;A_t (z)=0
\label{C6}
\ee
where $A_t (z)$ is a time component of the gauge $CB$-superfield.

One can choose a special $\tau$-gauge for the $SYM^1_6$-theory
\be
A_t (z)=0,\hspace{1cm} \partial_t \tau(z)=0 \label{C7}
\ee
The $A$-frame covariant derivative $\nabla_{12}$ commutes with $D^+_1$ and
 $D^+_2$. The simplest conserved quantities in the $\tau$-gauge can be
constructed as $\Lambda$-invariant functions of $h_{\s A}$, for example
\be
C^{\s++}(z,u)=\T (D^+_1h_{\s A}D^+_2h_{\s A}^{-1}),\hspace{1cm}
\partial_t C^{\s++}=0 \label{C8}
\ee

It is not difficult to built the conserved quantities invariant under the
$\tau$- and $\Lambda$-gauge transformations
\be
P^{ab\s(\pm\pm)}= \T W^{a\s\pm}\;W^{b\s\pm},\hspace{1cm}
\partial_{ab} P^{ab\s(\pm\pm)}=0 \label{C9}
\ee
where $W^{a\s\pm}$ are components of the on-shell superfield-strength
\be
W^{a+}= (D^{\s+3})^a\;A^{\s--},\;\;\;\;\;W^{a-}= \nabla^{\s--}\;
W^{a+} \label{C10}
\ee
\be
\nabla_{ab}\;W^{a\s\pm} = 0,\;\;\;\;\;\;\nabla_{ab}\;W^{a\s\pm}\;
W^{b\s\pm}\ = 0
\label{C11}
\ee
Note that the last equation is not valid off-shell.

The B\"acklund  transformations ($BT$) play an important role for
integrable theories as transformations in the spaces of solutions.
For the  $SDYM$ and $SDSYM$ solutions these transformations have been
 considered in Refs\cite{a24,a25}. We shall discuss $BT$ in the
 $HS$-formalism of $SYM^1_6$.

Let $A^{\s--}$ and $\hat{A}^{\s--}$ be two different solutions of the
$SYM^1_6$-system (\ref{B4}). Consider the corresponding bridges $h_{\s A}$
 and $\hat{h}_{\s\hat{A}}$. Then the B\"acklund transformation between
 these solutions has the following form:
\be
\hat{A}^{\s--} =D^+_a \hat{A}^{a\s(-3)}=B^{-1}\;A^{\s--}\;B +
B^{-1}\;\partial^{\s--}\;B \label{C12}
  \ee
  where the $B$-matrix can be written in terms of two bridges
  \be
  B(A,\hat{A})= h_{\s A}\; \hat{h}^{-1}_{\s\hat{A}}
 \label{C13}
  \ee

It is easy to derive the equations for the matrix $B$ in terms of the
background solution $A^{\s--},\;h_{\s A}$. Formally the new superfield
 variable
$\hat{A}^{\s--}$ has an independent $\hat{\Lambda}$ transformation, and it
 is 'invariant' under the $\Lambda$-transformation of a background
 superfield. Eq(\ref{C12}) can be treated as a harmonic equation for $B$
in terms of the background solution $A^{\s--},\;h_{\s A}$ and the second
 prepotential $\hat{A}^{a\s(-3)}$ . The analyticity equations (\ref{B4},
\ref{B10a}) for the second solution $\hat{A}^{\s--}$ produce the following
 $\Lambda$-covariant equations
\be
\nabla^{\s++}(A) \beta^+_a = 0,\;\;\;\;\;\nabla^{\s--}(A)\;
\nabla^{\s--}(A) \beta^+_a = 0 \label{C14}
\ee
where $\beta^+_a = D^+_a B\;B^{-1} $. Validity of these equations is
evident in the representation (\ref{C13})
 \be
 h^{-1}_{\s A}\;\beta^+_a\;h_{\s A} =
h^{-1}_{\s A}\;D^+_a\;h_{\s A} - \hat{h}^{-1}_{\s\hat{A}}\;D^+_a\;
\hat{h}_{\s\hat{A}}= u^+_i [ A^i_a(z) - \hat{A}^i_a(z) ] \label{C15}
 \ee

This representation is equivalent to the following form of the B\"acklund
transformation of the spinor $CB$ gauge superfield:
\be
\hat{A}^i_a(z)=A^i_a(z) + h^{-1}_{\s A}\;[u^i_-\beta^+_a\;-u^i_+
\nabla^{\s--}(A)\beta^+_a] h_{\s A}
\label{C15b}
\ee

The equations for $B$ are simplified in the case of infinitesimal
 B\"acklund transformations $B=I + \delta B$
\be
\delta B = D^+_{(a} \nabla^-_{b)}\; B^{ab} \label{C17}
\ee

The analyticity produces an additional restriction on $B^{ab}$.

\setcounter{equation}{0}
\section{\label{D} $V$-frame analysis of  $SU(2)$ solutions
in the simplest harmonic gauge}

$\;\;\;\;$ The $HS$-integrability interpretation allows us to analyze the
explicit constructions of the $SYM^1_6$-solutions by analogy with the
harmonic formalism of $SDYM$ \cite{a3,a4} or $SYM^6_3$ \cite{a12}. Let us
 go back to the $V$-frame and consider the case of the gauge group
$SU(2)$. We shall use a harmonic representation of the general $SU(2)$
 prepotential \cite{a12,a19,a20}
\be
V^{\s++} = (U^{\s+2})\; b^0 (\zeta,u) + (U^0)\;b^{\s(+2)} (\zeta,u) +
 (U^{\s-2})\;b^{\s(+4)} (\zeta,u) \label{D1}
\ee
where $b^{0},\;b^{\s(+2)},\;b^{\s(+4)}$ are arbitrary real analytic
 superfields
and $(U^{q})$ are matrix generators of the Lie algebra $SU(2)$ in a
harmonic representation
\be
(U^{\pm 2})^i_k = u^{\pm}_k \; u^{\pm i} ,\;\;\;\; (U^0 )^i_k =
      u^{-}_k\;u^{+i} + u^{+}_k\;u^{-i}  \label{D2}
\ee
An analogous representation of the prepotential was used as a special
Ansatz for instanton and monopole solutions in the harmonic formalism
\cite{a21,a22}.

Consider   the infinitesimal gauge transformations of the harmonic
components $ b^{\s(q)}$
\be
\delta b^{0}=\partial_A^{\s++}\Lambda^{\s(-2)} + 2\Lambda^{0} + 2b^0
 \Lambda^{0}- 2b^{\s(+2)}\;\Lambda^{\s(-2)}
                 \label{D3b}
\ee
\be
\delta b^{\s(+2)}=\partial_A^{\s++}\Lambda^0 + \Lambda^{\s(+2)} +
 b^{\s(+4)}\;\Lambda^{\s(-2)} - b^0 \Lambda^{\s(+2)}  \label{D3}
\ee
\be
\delta b^{\s(+4)}=\partial_A^{\s++}\Lambda^{\s(+2)} + 2 b^{\s(+2)}
\Lambda^{\s(+2)}
- 2 b^{\s(+4)}\;\Lambda^{0} \label{D4}
\ee
where $\Lambda^{\s(q)}$ are the real harmonic components of the analytic
 $SU(2)$-gauge matrix $\Lambda$. Remark that the $(\theta_+)^4$-component
 in (\ref{D4}) contains the term $\partial^{ab}\lambda_{ab}(x)$
with a total derivative of vector function from $\Lambda^{\s(+2)}$.

The simplest general gauge for  $SU(2)$-prepotential is
\be
V^{\s++}(b^0,\rho)=(U^{\s+2})\;b^0 (\zeta,u) + (U^{\s-2})\;(\theta_+)^4\;
\rho  \label{D5}
\ee
where $b^0$ is an arbitrary analytic function and $\rho$ is a constant
part of the trace  of the auxiliary scalar matrix field with $d=-2$ in
$b^{\s(+4)}$ that can be written as $D^{ik}_{ik}(x)=\rho + \partial^{ab}
 f_{ab}(x_{{\s A}})$. The $\rho\ne 0$ solutions characterize the
 phase of the $SYM^1_6$-theory with the spontaneous breaking of symmetry.

Stress that this $(b^0,\rho)$-gauge has the  residual
gauge invariance with $\Lambda^{\s(+2)}=0,\;\Lambda^0=const$ and an
arbitrary parameter $\Lambda^{\s(-2)}$. The additional condition
 $(\partial^{\s++})^5 b^0=0 $ fix the $\Lambda$-gauge and results in the
vanishing of harmonic components with isospin $T>4$ in $b^0$ \cite{a20}
\be
b^0(V_{iklm},u)=(D_4)^{iklm}\;V_{iklm} + 4 (u^+ u^-)_{ik}(D_4)^{lmn(i}\;
V^{k)}_{lmn} + \label{D6}
\ee
\[
(60/7)(u^{\s+2}u^{\s-2})_{ijkl}(D_4)^{mn(ij}V^{kl)}_{mn}+
(100/9)(u^{\s+3}u^{\s-3})_{i_1\cdots i_6}(D_4)^{n(i_1\cdots}
V^{\cdots i_6)}_{n} +
\]
\[
(50/9)(u^{\s+4}u^{\s-4})_{i_1\cdots i_8}(D_4)^{(i_1\cdots}\;
V^{\cdots i_8)}
\hspace{7cm}
\]
where an analogue of the Mezinchescu prepotential with $d=2$ \cite{Me} and
 the
irreducible symmetrical combinations of harmonics
$(u^{\s+q}u^{\s-q})_{i_1\cdots i_{2q}}$ are used. The analyticity of this
 representation follows from  the identity (\ref{A3b}).

The phase of $SYM^1_6$ and $SYM^2_4$ with $\rho=0$ was considered in
Refs\cite{a12,a19,a20}.
The $HZC$-equation (\ref{A13}) has the following solution in the
 $(b^0,0)$-gauge
\be
A^{\s--}(b^0,0) = (U^{\s+2})\;a^{\s(-4)}_0 + (U^0)\;a^{\s(-2)}_0 +
  (U^{\s-2})\;a^{\s(0)}_0                \label{D7}
	      \ee
where $a^{\s(q)}_0$ are   harmonic-quadrature functions of the
prepotential $b^0$
\be
a^{\s(0)}_0  = \frac{b(z)}{1+b(z)}\;,\;\;\;\;\;b(z)=\int du b^0 (z,u)
 \label{D8}
\ee
\be
a^{\s(-2)}_0(z,u)=\int du_1 \frac{(u^- u^+_1 )}{(u^+ u^+_1 )}\;\frac{b^0
 (z,u_1) - b(z)}{1+b(z)} \label{D9}
\ee
\be
a^{\s(-4)}_0(z,u)=[1+b(z)]\;[\partial^{\s--}\;a^{\s(-2)}_0 -
a^{\s(-2)}_0\;a^{\s(-2)}_0] \label{D10}
\ee
Note that this solution has a singular point $b(z)=-1$.

The classical action of $SYM^1_6$ in the $(b^0,0)$-gauge has the following
 form \cite{a19,a20}:
\be
S(b)=\int d^{14}z [\mbox{ln}(1+b(z))-b(z)] \label{D10b}
\ee
where $b(z)=(D_4)^{iklm}\;V_{iklm}(z)$ is a constrained potential.

The $SYM^1_6$-equation of motion in the $(b^0,0)$-gauge has  only one
 independent  component
\be
 (D^+ )^4 \left[\frac{b(z)}{1+b(z)}\right] = 0
 \label{D11}
\ee

A spinor part of the gauge $CB$-superfield   can be written in terms
of the single superfield $b(z)$ \cite{a20}
\be
\left[A^l_a(z)\right]^k_i = \frac{1}{1+b(z)}\left[\delta^l_i D^k_a b(z) -
(1/2)\delta_i^k D^l_a b(z)\right] \label{D12}
\ee
Note that the $SYM^1_6$-constraints (\ref{A4}) in this representation
follow from the identity
\be
 (D_2)_{ab}b(z)=0  \label{D11b}
\ee

The harmonic equations (\ref{A8}) and  (\ref{A13}) with the prepotential
 $V^{\s++}(b^0,\rho)$ (\ref{D5}) can be integrated in quadratures.
The integration procedure uses a nilpotency of the term
$\rho(\theta_+)^4$.

Eq(\ref{A13}) has the following harmonic components in the
 $(b^0,\rho)$-gauge:
\be
\partial^{\s++} a^{\s(-4)}_\rho + 2(1+b^0)a^{\s(-2)}_\rho -
\partial^{\s--}b^0 =0 \label{D12b}
\ee
\be
\partial^{\s++} a^{\s(-2)}_\rho +(1+b^0)a^{\s(0)}_\rho -b^0 -
\rho (\theta_+)^4 a^{\s(-4)}_\rho=0 \label{D12c}
\ee
\be
\partial^{\s++} a^{\s(0)}\rho -4\rho\theta^a_-(\theta^{+3})_a -
2\rho(\theta_+)^4 a^{\s(-2)}_\rho=0 \label{D12d}
\ee
Note that it is convenient to analyze harmonic equations in the central
 coordinates $z,u$.

Consider the harmonic equation for $a^{\s(0)}_\rho $ which follows
from these equations
\be
(\partial^{\s++})^2 a^{\s(0)}_\rho=2\rho(\theta_+)^4[2+b^0 -
a^{\s(0)}_\rho - b^0 a^{\s(0)}_\rho ] \label{D13}
\ee

Using  (\ref{D8}) as a zero approximation one can obtain an
exact solution for $a^{\s(0)}_{\rho} $ by two iterations and then the
other harmonic components can be calculated.

The classical action in the $(b^0,\rho)$-gauge has the following form:
\be
S(b^0,\rho)=\int d^{14}z\, du\; b^0 \int^1_0 ds\; a^{\s(0)}(sb^0,\rho)
\ee
where $s$ is an auxiliary parameter.

\setcounter{equation}{0}
\section{ \label{E} The $A$-frame analysis of $SU(2)$-solutions}

\hspace{5mm}Now we shall discuss properties of the $SU(2)$-solution in the
alternative $A$-frame. The first step of this approach is a solution of
  harmonic equations in the representation (\ref{B2}) and then the
 dynamical analyticity equation should be used.

Consider the harmonic $(U^q)$-components of the $AZC$-equation (\ref{B3})
\be
D^-_a a^{\s(0)} + \partial^{\s--}D^+_a a^{\s(0)}+2D^+_a a^{\s(-2)} +
2 a^{\s(-2)}D^+_a a^{\s(0)}-2a^{\s(0)}D^+_a a^{\s(-2)}=0  \label{F0}
 \end{equation}
\be
D^-_a a^{\s(-2)} + \partial^{\s--}D^+_a a^{\s(-2)}+ D^+_a a^{\s(-4)} +
 a^{\s(-4)}D^+_a a^{\s(0)} - a^{\s(0)}D^+_a a^{\s(-4)}=0  \label{F0b}
\end{equation}
\be
D^-_a a^{\s(-4)} + \partial^{\s--}D^+_a a^{\s(-4)}+
 2a^{\s(-4)}D^+_a a^{\s(-2)} - 2a^{\s(-2)}D^+_a a^{\s(-4)}=0  \label{F0c}
\end{equation}

These equations are equivalent to the dynamical equations
 $D^+_a b^{\s(q)}(A)=0$ for the harmonic components of $V^{\s++}$
(\ref{D1}) in the $A$-frame.

The analyticity equations imply the following condition
\be
\nabla^{\s++}W^{+a}=\nabla^{\s++}(D^{\s+3})^a A^{\s--}=0 \label{F1}
\end{equation}
 producing the relations between harmonic components of $W^{+a}$.
Remark that the additional conditions $D^+_a a^{\s(0)}=0$ or
$(D^{\s+3})^a a^{\s(-2)}=0$ correspond to pure gauge solutions
$W^{+a}=0$.

By analogy with (\ref{F1}) one can obtain the general relations between
the harmonic components of Eq(\ref{B1}):
\be
(D^+)^4 a^{\s(-4)}=0 \Leftrightarrow (D^+)^4 a^{\s(0,-2)}=0 \label{F2}
\end{equation}
The on-shell dependence of the superfields $a^{\s(q)}$ allow us to
simplify the $SYM^1_6$-equations.

Now the convenient 'hybrid' choice of the field variables will be
considered.
Let $a^{\s(0)},a^{\s(-2)}$, $b^{\s(+2)}$ and $b^{\s(+4)}$ be the
 independent variables and $b^{0}$ and $a^{\s(-4)}$
be treated as the functions of these variables. We can use the gauge
(\ref{D5}) and Eqs(\ref{D12b}-\ref{D12d}) in this frame, too.

 Using Eq(\ref{D12c}) one can obtain the relation for the dependent
function of the hybrid frame
\be
b^0(A)=\frac{1}{1- a^{\s(0)}_\rho }[\partial^{\s++} a^{\s(-2)}_\rho +
a^{\s(0)}_\rho-\rho
(\theta_+)^4  a^{\s(-4)}_\rho ] \label{F3}
\end{equation}

The analyticity condition $D^+_a b^0(A)=0$ is a single dynamical equation
in this approach. It should be stressed that this equation describe the
general $SU(2)$ solution.

Consider a solution of the harmonic bridge equation (\ref{B13})
for the case $\rho=0$
\be
h_{\s A}=\mbox{exp}[(1/2)(U^0)\mbox{ln}(1-a^{\s(0)}_0)][1-(U^{\s+2})
a^{\s(-2)}_0 ] \label{F3b}
\end{equation}
This solution has only one singular point $a^{\s(0)}_0=1$. More general
solution can contain additional singularities. An arbitrariness in the
bridge solution is connected with the gauge freedom of Eq(\ref{B13}).
Eq(\ref{F3b}) produces a relation for
$a^{\s(-4)}_0$ analogous to (\ref{D10}).

 The polynomial form of the corresponding dynamical equation is
\be
(1 + \partial^{\s++}a^{\s(-2)}_0)\;D^+_a\;a^{\s(0)}_0 + (1-a^{\s(0)}_0 )\;
D^+_a\;\partial^{\s++}a^{(\s-2)}_0 = 0 \label{F4}
\end{equation}
\be
a^{\s(0)}_0 (z)=(D_2)_{ab}\;A^{ab}(z),\;\;\;\;a^{\s(-2)}_0(z,u)=D^+_a\;
A^{a\s(-3)}(z,u) \label{F5}
\end{equation}

Remark that this one-component equation is covariant under the residual
gauge transformations of the $(b^0,0)$-gauge . The consistency condition
for this equation follows from the restriction (\ref{D11b})
\be
(D_2)_{ab}\int du b^0(z,u)=(D_2)_{ab}\left[\frac{a^{\s(0)}_0}
{1-a^{\s(0)}_0} \right]=0 \label{F5b}
\end{equation}
One can try to solve these equations in superfields or in components and
 then
 use the $b^0$-solution for the construction of the bridge to the central
basis.

Thus, the $SYM^1_6$-system reduces to Eqs(\ref{D11}) or (\ref{F4}) in
the $(b^0,0)$-gauge. This reduction simplifies significantly the
initial $SYM^1_6$-system and gives the hope to obtain the explicit
solutions of this problem.

\setcounter{equation}{0}
\section{ Conclusion }

$\;\;\;\;$The harmonic-superspace integrability of $SYM^1_6$-theory
guarantees the analogous property of its $N=2, D=4$ subsystem $SYM^2_4$.
Consider the representation (\ref{B2}) in the Euclidean version of
$SYM^2_4$
\be
A^{\s--}(z,u)=D^+_\alpha \;A^{\alpha\s(-3)} + \bar{D}^+_{\dot{\alpha}} \;
\bar{A}^{\dot{\alpha}\s(-3)} \label{E1}
\end{equation}
where two-component spinors are used.

The case $A^{\alpha\s(-3)}=0$ corresponds to the general self-dual
solution of $SYM^2_4$
\be
W(A)=(\bar{D}^+)^2\;A^{\s--} = 0 \label{E2}
\end{equation}
The self-dual prepotential $\bar{A}^{\dot{\alpha}\s(-3)}$ satisfies also
the nonlinear $AZC$-equation (\ref{B4}).

Note that $SYM^2_4$-equations in $HS$ are covariant under the discrete
transformation
\be
\theta^\alpha_i \leftrightarrow \bar{\theta}^{\dot{\alpha}}_i\hspace{1cm}
A^{\alpha\s(-3)} \leftrightarrow \bar{A}^{\dot{\alpha}\s(-3)}
\label{E3}
\end{equation}
that is a residual form of the Lorentz transformation in $D=6$. This
discrete transformation corresponds to the duality transformation between
 self-dual and anti-self-dual solutions. Note that other discrete
 transformations exist in the 1-st order formalism of $SYM^2_4$ with the
 independent field-strengthes
$F^\alpha_\beta, F^{\dot{\alpha}}_{\dot{\beta}}$ and
 $F^\alpha_{\dot{\beta}}$.

It is interesting to discuss possible alternative forms of the
$SYM^2_4$-action. A simple possibility is the action of a gauge-invariant
 harmonic interaction of the {\it independent unconstrained}  harmonic
superfields $h, V^{\s++},A^{a\s(-3)} $ and  $\;L^{a\s(-3)}$
\be
S(h,V,A,L)=\int dz\;du \mbox{Tr} [(D^+_a A^{a\s(-3)} +\partial^{\s--}h
 h^{-1})(V^{\s++}\;+ \partial^{\s++}h h^{-1})\;+\;L^{a\s(-3)} D^+_a
 \partial^{\s++}h h^{-1}]  \label{E4}
\ee
where $a=\alpha,{\dot{\beta}}$. The first part of $S$ is a product of
 two covariant terms in the framework of the above-mentioned
$\Lambda$-transformations of $h,V,A$, and a choice of the
$L$-transformation is evident.

This $\sigma$-model-type action produces the standard on-shell analyticity
 and other dynamical equations for the bridge $h$, however, it contains
additional degrees of freedom. Specific features of $SYM^2_4$-solutions
will be discussed elsewhere.

It seems natural that the effective quantum action of $SYM^2_4$
 \cite{SW} can be rewritten in terms of $N=2$ superfields . Note
that the simplest harmonic gauge for the gauge group $SU(3)$ contains
analytic components $b^0_3$ and $b^{(+2)}_8$ corresponding to the Cartan
 generators of $SU(3)$ \cite{a19,a20}.  Analogous harmonic gauges can be
 found for any gauge group.

The  integrable theory $SYM^3_4$ can be described in the framework of
 $SYM^2_4$ with the special hypermultiplet interactions \cite{a2}. An
analogous construction exists for the integrable $SYM^2_6$-theory in
terms of $HS^1_6$-superfields. It seems natural to consider the
$A$-frame $HS$-equations of  more general interacting
 $SYM$-supergravity-matter systems. Any $HS$-integrable system can be
 reduced to the dynamical analyticity conditions and some solvable
 linear constraints. This formulation may help to build the explicit
 classical solutions and to study quantum solutions.

The author would like to thank cordially  V.I. Ogievetsky, E.A. Ivanov and
C. Devchand for stimulating discussions and critical remarks and
A.A. Kapustnikov, A.D. Popov and K.S. Stelle  for discussions.
I am grateful to A.T.Filippov and the administration of
 LTP JINR for support
and hospitality. This work is partially supported by ISF-grant RUA000 and
INTAS-grant 93-127.

\end{document}